\documentstyle[11pt]{article}
\begin{document}
{\setlength{\oddsidemargin}{1.2in}
\setlength{\evensidemargin}{1.2in} } \baselineskip 0.55cm
\begin{center}
{ {\bf Quantum radiation of Kerr black hole in de Sitter background}}
\end{center}
\begin{center}
${\rm T.\ Ibungochouba\ Singh^{^1}, Y.\ Kenedy \ Meitei^{1}, I.\ Ablu \ Meitei^{2}, K.\ Yugindro\ Singh^{3}}$
\end{center}
\begin{center}
1. Mathematics Department, Manipur University, Canchipur, 795003, Manipur(India)\\
2. Physics Department, Modern College, Imphal, Manipur (India)\\
3. Physics Department, Manipur University, Canchipur, 795003, Manipur (India)\\
1.ibungochouba@rediffmail.com\\
\end{center}
\date{}

\begin{abstract}
In this paper, we investigate the tunneling of vector boson particles across the event horizon of Kerr-de Sitter black hole by using Hamilton-Jacobi ansatz to Proca equation and WKB approximation. The surface gravity of KdS black hole has been recovered by using direct calculation and Proca equation. These two different methods give the same Hawking temperature at the event horizon.

\end{abstract}

1. {\bf Introduction}

The thermal radiation of black hole was discovered by  Hawking [1,2] using the techniques of quantum field theory in curved spacetime. The study of thermal and non-thermal radiation properties of black holes is an important aspect in black hole physics. In the last few decades, much attention has been paid to investigate the thermal and non-thermal radiations of scalar particles and Dirac particles in many different types of black holes. Refs. [3-5] have shown that the entropy of a black hole is equal to one fourth times the area of the event horizon of the black hole. Ref. [6] introduced the tortoise coordinate transformation in which the gravitational field is independent of time. The particle emission arises directly from a quantum mechanical barrier penetration across the black hole event horizon. Sannan [7] studied the probability distributions of both bosons and fermions emitted by black hole.

Refs. [8-10] proposed the Hawking radiation as a semiclassical quantum tunneling phenomenon where a particle moves in a dynamical geometry and derived the tunneling potential barrier created by the outgoing particle. Assuming energy conservation and unfixed background space time, the radial null geodesic of the particle and imaginary part of the action can be obtained by using the relativistic Hamilton-Jacobi equation and WKB approximation. Chattterjee et al. [11] also studied the Hawking radiation as tunneling by Hamilton-Jacobi method for non-rotating and rotating black holes. Using different non-singular coordinate systems, they not only confirmed the quantum emission from black hole but also revealed the new phenomenon of absorbtion into white holes by quantum mechanical tunneling. Following their method much works have been done [12-16]. Banerjee and Majhi [17] investigated the Hawking radiation as tunneling by Hamilton-Jacobi method beyond the semiclassical approximation. The correction to Bekenstein-Hawking area law from the modified Hawking temperature is obtained by using the laws of black hole mechanics.  Using the method devised by Banerjee and Majhi, many fruitful results have been obtained in [18-21].

Ghosh  at el. [22, 23] discussed the scaling of black hole entropy in loop quantum gravity. They showed that the black hole entropy differs from the low energy expected value $A/4$ in the deep Planckian region. The black hole entropy based on the near horizon symmetries of black hole space time has been discussed and applying the Cardy formula and the central charge, the Bekenstein-Hawking entropy was recovered in [24]. Ref. [25] derived the Hawking radiation for the near-extremal Reissner-Nordstorm and Kerr black holes. A dynamically evolving black hole can be assigned a Hawking temperature and the Hawking flux was also recovered in [26].

Recently, Refs. [27, 28] proposed Hawking radiation of vector particles from black holes by using the Hamilton-Jacobi ansatz to Proca equation and WKB approximation. For the Schwarzschild background geometry the emission temperature is the same as Hawking temperature corresponding to scalar particle emission. Following their works, different authors have investigated the quantum tunneling of massive vector particles by using the Hamilton-Jacobi ansatz to Proca equation and WKB approximation in different types of spacetime such as three dimensional rotating hairy black  hole [29], Schwarzschild black hole[30], Kerr and Kerr-Newman black hole [31], Lorentzian Wormhole in 3+1 dimensions [32], Warped ${\rm AdS_{3}}$ black  hole [33],  Kerr-Newman black hole in dragging coordinate, retarded coordinate and Painleve coordinate [34]. Chen et al. [35] investigated the Hawking radiation as tunneling from rotating black holes in de Sitter background by using Dirac equation. The Hawking temperatures at the black hole event horizon and cosmological horizon in de Sitter background have been recovered.

The non-zero value of the cosmological constant, $\Omega_{\Lambda}$ signifies the existence of a form of energy density with negative pressure which is believed to be responsible for the accelerated expansion of the universe [36-38].  It is an interesting aspect as to whether the cosmological constant has any effect on the Hawking radiation. In this paper, we aim to study the Hawking radiation of massive vector boson particles from Kerr-de Sitter black hole. The tunneling of the vector particles from the event horizon of Kerr-de Sitter black hole is studied by using the Hamilton-Jacobi ansatz to the Proca equation and WKB approximation.

The paper is organized as follows: in section 2, we derive the Hawking temperature of Kerr-de Sitter black hole in the dragging coordinate system by using the Hamilton-Jacobi ansatz to Proca equation and WKB approximation; in section 3, the result and discussion is given; in the last section some conclusions are given in the last section.

2. {\bf Kerr-de Sitter black hole (KdS)}.

 The  metric of KdS black hole in the Boyer-Lindguist type coordinates $(t, r, \theta, \phi)$ with cosmological constant $\Lambda$ is given by [39]
\begin{eqnarray}
ds^2&=&-\frac{A[\Delta-\Delta_\theta a^2\sin^2\theta]}{R^2}dt^2+\frac{R^2}{\Delta}dr^2+\frac{R^2}{\Delta_\theta}d\theta^2
-\frac{2aA[\Delta_\theta(r^2+a^2)-\Delta]\sin^2\theta}{R^2}dtd\phi\cr&&
+\frac{A\sin^2\theta[\Delta_\theta(r^2+a^2)^2-\Delta a^2\sin^2\theta}{R^2}] d\phi^2,
\end{eqnarray}
where
\begin{eqnarray}
&&R^2=r^2+a^2\cos^2\theta,\,\,\,\,\, A=[1+\frac{1}{3}\Lambda a^2]^{-2},\cr
&&\Delta_\theta=1+\frac{1}{3}\Lambda a^2\cos^2\theta,\cr
&&\Delta=(r^2+a^2)\Big(1-\frac{1}{3}\Lambda r^2\Big)-2Mr.
\end{eqnarray}
Here M, $\Lambda$ and $a$ are the mass, cosmological constant and rotational parameter of KdS black hole respectively.
The KdS black hole has apparent singularities at the value of $r$ for which $\Delta=0$, that is
\begin{equation}
r^2+a^2-2Mr-\frac{1}{3}\Lambda r^2(r^2+a^2)=0.
\end{equation}
 Eq. (3) has four real roots for $\Lambda>0$. The four roots of Eq. (3) say $r_c$, $r_h$, $r_1$ and  $r_-$ $(r_c>r_h>r_1>0>r_-$) will satisfy the following equation:
\begin{equation}
(r-r_c)(r-r_h)(r-r_1)(r-r_-)=-\frac{3}{\Lambda}[r^2+a^2-2Mr-\frac{1}{3}\Lambda r^2(r^2+a^2)].
\end{equation}
The biggest root $r_c$ denotes the location of the cosmological horizon of the black hole;  $r_1$ indicates the Cauchy horizon of the black hole and $r_h$ is the location of the black hole event horizon and also the smallest negative root, $r_-$ represents the another cosmological horizon on the other side of the ring singularity at $r=0$ and another at infinity [40]. We factorize $\Delta$ into the following form:
\begin{eqnarray}
\Delta=(r-r_h)\Delta'(r_h)
\end{eqnarray}
where ${\frac{d\Delta}{dr}}{\mid}_{r=r_h}=\Delta'(r_h)$. The metric (1) has a singularity at the radius of the event horizon.
 By using dragging coordinate transformation, $\frac{d\phi}{dt}=-\frac{g_{14}}{g_{44}}$ [41], the line element of the KdS black hole can be expressed as:
\begin{eqnarray}
ds^2=\hat{g}_{11}dt^2_k+\frac{R^2}{\Delta}dr^2+\frac{R^2}{\Delta_\theta}d\theta^2,
\end{eqnarray}
where $\hat{g}_{11}=-\frac{A\Delta \Delta_\theta R^2}{[\Delta_\theta(r^2+a^2)^2-\Delta a^2\sin^2\theta]}$. Eq. (6) denotes a 3-dimensional hypersurface in 4-dimensional KdS black hole. The dragging velocity at the black hole horizon $r=r_h$ is given by
\begin{eqnarray}
\Omega_h=\frac{a}{r^2_h+a^2}.
\end{eqnarray}
The surface gravity at $r=r_h$ is given by [42]
\begin{eqnarray}
\kappa&=&\lim_{\hat{g}_{11}\rightarrow 0}\Big(-\frac{1}{2}\sqrt{\frac{-g^{22}}{\hat{g}_{11}}}\frac{d\hat{g}_{11}}{dr}\Big)=
\frac{\sqrt{A}(r_h-M-\frac{2\Lambda r^3_h}{3}-\frac{\Lambda r_h a^2}{3})}{(r^2_h+a^2)}.
\end{eqnarray}
The Hawking temperature of the black hole is given by
\begin{eqnarray}
T&=&\frac{1}{2\pi}\frac{\sqrt{A}(r_h-M-\frac{2\Lambda r^3_h}{3}-\frac{\Lambda r_h a^2}{3})}{(r^2_h+a^2)}.
\end{eqnarray}
We will study tunneling of massive vector particles of KdS black hole. Within the semi-classical approximation, the wave function $\Psi$ satisfies the Proca equation as
\begin{equation}
\frac{1}{\sqrt{-g}}\partial_{_{a}}(\sqrt{-g}\Psi^{ab})+\frac{m^2}{{\hbar}^{2}}\Psi^b=0,
\end{equation}
where $\Psi_{ab}=\partial_{a}\Psi_{b}-\partial_{b}\Psi_{a}$ and $\Psi^{ab}$ is an anti-symmetric tensor. From Eqs. (6) and (10), we obtain the components of wave function $\Psi$ as follows:

\begin{eqnarray}
\Psi^{0}&=&-\frac{B}{A\Delta\Delta_\theta R^2}\Psi_{0},\,\,\, \Psi^{1}=\frac{\Delta}{R^2}\Psi_{1},\,\,\,\,\,\,\Psi^{2}=\frac{\Delta_\theta}{R^2}\Psi_{2}\cr
\Psi^{01}&=&-\frac{B}{A\Delta_\theta R^2R^2}\Psi_{01},\,\,\,\,\,\,\, \Psi^{02}=-\frac{B}{A\Delta R^2R^2}\Psi_{02},\,\,\,\Psi^{12}=\frac{\Delta \Delta_\theta}{R^2R^2}\Psi_{12},
\end{eqnarray}
 where $B=\Delta_\theta(r^2+a^2)^2-\Delta a^2\sin^2\theta$. Then the Proca equations are reduced to
\begin{eqnarray}
\frac{\partial}{\partial r}(\sqrt{-g}\Psi^{01})+\frac{\partial}{\partial \theta}(\sqrt{-g}\Psi^{02})+(\hbar^{2})^{-1}m^2\sqrt{-g}\Psi^0=0,\cr
\frac{\partial}{\partial t}(\sqrt{-g}\Psi^{10})+\frac{\partial}{\partial \theta}(\sqrt{-g}\Psi^{12})+(\hbar^{2})^{-1}m^2\sqrt{-g}\Psi^1=0,\cr
\frac{\partial}{\partial t}(\sqrt{-g}\Psi^{20})+\frac{\partial}{\partial r}(\sqrt{-g}\Psi^{21})+(\hbar^{2})^{-1}m^2\sqrt{-g}\Psi^2=0
\end{eqnarray}
The vector function can be defined as
\begin{eqnarray}
\Psi_a=(c_0, c_1, c_2) {\rm exp}\Big [\frac{i}{\hbar}S(t_{k},r,\theta)\Big].
\end{eqnarray}
Using the WKB approximation, the action can be written as [17]
\begin{eqnarray}
S(t_{k}, r, \theta)=S_{0}(t_{k}, r, \theta)+\hbar S_{1}(t_{k}, r, \theta)+\hbar^{2} S_{2}(t_{k}, r, \theta)....
\end{eqnarray}
 Using Eqs. (11), (13) and (14)  in Eq. (12), neglecting the higher order terms of $o(\hbar)$, the resulting equations can be obtained as follows:
\begin{eqnarray}
&&c_2\Delta_\theta\frac{\partial S_0}{\partial \theta}\frac{\partial S_0}{\partial t_{k}}-c_0\Delta_\theta\Big(\frac{\partial S_0}{\partial \theta}\Big)^2+\Delta c_1\frac{\partial S_0}{\partial r}\frac{\partial S_0}{\partial t_{k}}-\triangle c_0\Big(\frac{\partial S_0}{\partial r}\Big)^2-m^2R^2c_0=0,\cr\cr
&&c_0B\frac{\partial S_0}{\partial t_{k}}\frac{\partial S_0}{\partial r}-B c_1\Big(\frac{\partial S_0}{\partial t_{k}}\Big)^2-A\Delta^{2}_{\theta}\Delta  c_2\frac{\partial S_0}{\partial \theta}\frac{\partial S_0}{\partial r}+A\Delta^{2}_{\theta}\Delta c_1(\frac{\partial S_0}{\partial \theta})^2\cr&&+A\Delta_\theta \Delta m^2R^2c_1=0,\cr\cr
&&c_0B(\frac{\partial S_0}{\partial t_{k}})(\frac{\partial S_0}{\partial \theta})-B c_2\Big(\frac{\partial S_0}{\partial t_{k}}\Big)^2-\Delta^2 A\Delta_\theta  c_1(\frac{\partial S_0}{\partial r})(\frac{\partial S_0}{\partial \theta})+A\Delta_\theta\Delta^2 c_2\Big(\frac{\partial S_0}{\partial r}\Big)^2\cr&&+A\Delta_\theta\Delta m^2R^2c_2=0.
\end{eqnarray}
It would be very difficult to find the action $S_0$ directly from Eqs. (15). We assume the solution as [21]
\begin{eqnarray}
S_0=-\omega t_{k}+W(r)+K(\theta)+\zeta,
\end{eqnarray}
where $\omega$ is the energy of the vector particle and $\zeta$ is a complex constant. Putting Eq. (16) into Eq. (15), we obtain the matrix equation
\begin{eqnarray}
\Lambda(c_0, c_1, c_2)^T=0,
\end{eqnarray}
where $\Lambda$ is a 3x3 matrix and superscript T means the transition to the transposed vector. Then, the components of $\Lambda$ matrix are given below:
\begin{eqnarray}
&&\Lambda_{00}=\Delta_\theta K^2_\theta+\Delta  W^2_r+m^2R^2,\,\,\,\,\Lambda_{01}=\Delta \omega W_{r},\,\,\,\Lambda_{02}=\Delta_\theta K_{\theta}\omega,\cr
&&\Lambda_{10}=-\omega BW_{r},\,\,\,\,\Lambda_{11}=-\omega^2 B+A\Delta^{2}_\theta \Delta K^2_{\theta}+\Delta m^2R^2A\Delta_\theta,\cr
&&\Lambda_{12}=-A\Delta\Delta^{2}_\theta K_{\theta}W_{r},\,\,\,\,\Lambda_{20}=-B\omega K_{\theta},\,\,\,\,\,\,\Lambda_{21}=-A\Delta_\theta\Delta^2 W_{r}K_{\theta},\cr
&&\Lambda_{22}=-B\omega^2+A\Delta_\theta\Delta^2W^2_{r}+A\Delta_\theta\Delta R^2m^2,
\end{eqnarray}
where $W_{r}=\frac{\partial W}{\partial r}$ and $K_{\theta}=\frac{\partial K}{\partial \theta}$.
Eq. (18) is a homogeneous system of linear equations and it admits non-trivial solution if and only if $\rm{det}\Lambda(c_0, c_1, c_2)=0$.
i.e.
\begin{eqnarray}
&&(A\Delta_\theta\Delta m^2R^2-B\omega^2+A\Delta_\theta\Delta^2W^2_{r}+A\Delta^{2}_\theta\Delta K^2_{\theta})[(\Delta m^2R^2A\Delta_\theta-B\omega^2)(\Delta_\theta K^2_{\theta}\cr&&+m^2R^2+\Delta W^2_{r})+B\Delta_\theta K^2_{\theta}\omega^2+\Delta B \omega^2 W^2_{r}]=0.
\end{eqnarray}
On integration, we obtain
\begin{eqnarray}
W_{\pm}=\pm\int\sqrt{\frac{(B\omega^2-A\Delta_\theta\Delta m^2 R^2)(\Delta_\theta K^2_\theta+m^2R^2)-BK^2_\theta\omega^2}{A\Delta_\theta\Delta^2m^2R^2}}dr,
\end{eqnarray}
where $W_{+}$ corresponds to outgoing spin-1 particle (moving away from the black hole) and $W_{-}$ corresponds to the ingoing (moving towards the black hole) spin-1 particle. From Eq. (20), we see that there is a pole at the black hole horizon $r=r_{h}$ and imaginary part of the action can be derived from the pole. Using Feynman prescription and completing the integral, we get
\begin{eqnarray}
W_{+}\equiv\frac{i\pi\omega(r^2_h+a^2)}{2\sqrt{A}(r_h-M-\frac{2\Lambda r^3_h}{3}-\frac{\Lambda r_h a^2}{3})}
\end{eqnarray}
 The probabilities of the particles crossing the black hole horizon $r=r_{h}$ are given by
\begin{eqnarray}
\Gamma_{\rm emission}=exp(-2\rm Im S_0)=exp[-2(Im W_{+}+Im \zeta)]
\end{eqnarray}
and
\begin{eqnarray}
\Gamma_{\rm absorbtion}=exp(-2\rm Im S_0)=exp[-2(Im W_{-}+Im \zeta)].
\end{eqnarray}
There exists $100\%$ chance for the ingoing spin--1 particle to enter the black hole according to the semiclassical approximation. This shows that $\rm Im \zeta=-\rm ImW_-$. Since $W_{-}=-W_{+}$, the probability of outgoing particles can be obtained as
\begin{eqnarray}
 \Gamma_{\rm emission}=exp(-4 \rm Im W_{+})
\end{eqnarray}
Then, the tunneling rate can be expressed as follows
\begin{eqnarray}
\Gamma_{rate}=\frac{\Gamma_{\rm emission}}{\Gamma_{\rm absorbtion}}=exp(-4\rm Im W_{+})=exp\Big[\frac{-2\omega\pi(r^2_h+a^2)}{\sqrt{A}(r_h-M-\frac{2\Lambda r^3_h}{3}-\frac{\Lambda r_h a^2}{3})}\Big],
\end{eqnarray}
which is equivalent to the Boltzmann factor; $exp(-\omega_{net}\beta)$, where $\beta$ is the inverse temperature of the black hole.
Thus, the Hawking temperature is recovered at the black hole horizon as
\begin{eqnarray}
T&=&\frac{1}{2\pi}\frac{\sqrt{A}(r_h-M-\frac{2\Lambda r^3_h}{3}-\frac{\Lambda r_h a^2}{3})}{(r^2_h+a^2)},
\end{eqnarray}
which is the same as the Hawking temperature derived from the exact calculation as given in Eq. (9).

3. {\bf Result and Discussion}.
We discuss the Hawking temperature of Kerr-de Sitter black hole by using direct calculation and Hamilton-Jacobi ansatz to Proca equation. For determining the Hawking temperature as given by using the two methods, the line element given in Eq. (1) which has singularity at the event horizon is transformed using the dragging coordinate transformation to a new coordinate system. From Eqs. (9) and (26), it is observed that the two methods give the same Hawking temperature.

4. {\bf Conclusion}.

The Hawking radiation of massive vector particles tunneling across the event horizon of stationary KdS black hole has been discussed by applying Hamilton-Jacobi ansatz to Proca equation. In the derivation of tunneling rate within the framework of WKB approximation, the main aim is to calculate the imaginary part of the action which gives the tunneling rate of the vector particles that cross the event horizon. When we compare it with the Boltzmann factor the expected Hawking temperature of KdS black hole at the black hole horizon is recovered and is shown to be consistent with black hole universality.

{\bf Acknowledgements} : Kenedy acknowledges the CSIR for providing financial support.

\end{document}